\documentclass[amssymb,amsmath,prd,twocolumn,preprintnumbers,superscriptaddress,nofootinbib]{revtex4-1}
\usepackage[utf8]{inputenc}
\usepackage{color}
\usepackage{amsmath,bm, physics}
\usepackage{amssymb, mathtools}
\usepackage{hyperref}
\usepackage{cleveref}
\usepackage{graphicx,xcolor}
\usepackage{dcolumn}
\usepackage{bm}
\usepackage{ulem,cancel,soul}

\newcommand{\OmegaGW}{\Omega_{\rm GW}}
\newcommand{\barOmegaGW}{\bar\Omega_{\rm GW}}

\newcommand{\zmax}{z_{\rm max}}
\newcommand{\Tobs}{T_{\rm obs}}

\begin{document}

\title{Projections of the uncertainty on the compact binary population background using {\tt popstock}}
\author{Arianna I. Renzini}
\email{arianna.renzini@unimib.it}
\affiliation{Dipartimento di Fisica “G. Occhialini”, Universit\`a  degli Studi di Milano-Bicocca, Piazza della Scienza 3, 20126 Milano, Italy}
\affiliation{INFN, Sezione di Milano-Bicocca, Piazza della Scienza 3, 20126 Milano, Italy}

\author{Jacob Golomb}
\affiliation{Department of Physics, California Institute of Technology, Pasadena, California 91125, USA}
\affiliation{LIGO  Laboratory,  California  Institute  of  Technology,  Pasadena,  California  91125,  USA}

\begin{abstract}
    The LIGO-Virgo-KAGRA collaboration has announced the detection of almost 100 binary black holes so far, which have been used in several studies to infer the features of the underlying binary black hole population. From these, it is possible to predict the overall gravitational-wave (GW) fractional energy density contributed by black holes throughout the Universe, and thus estimate the gravitational-wave background (GWB) spectrum emitted in the current GW detector band. These predictions are fundamental in our forecasts for background detection and characterization, with both present and future instruments. The uncertainties in the inferred population strongly impact the predicted energy spectrum, and in this paper we present a new, flexible method to quickly calculate the energy spectrum for varying black hole population features such as the mass spectrum and redshift distribution. We implement this method in an open-access package, {\tt popstock}, and extensively test its capabilities. Using {\tt popstock}, we investigate how uncertainties in these distributions impact our detection capabilities and present several caveats for background estimation. In particular, we find that the standard assumption that the background signal follows a $2/3$ power-law at low frequencies is both waveform and mass-model dependent, and that the signal power-law is likely shallower than previously modelled, given the current waveform and population knowledge.
\end{abstract}

\maketitle

\section{Introduction}

The Laser Interferometer Gravitational-wave Observatory (LIGO)~\cite{LIGO}, Virgo~\cite{Virgo}, and KAGRA~\cite{KAGRA} detectors are progressively uncovering the features of the population of merging stellar-mass binary black holes in our Universe~\cite{pops-GWTC2,pops-GWTC3}. %
As observing runs become more and more sensitive, the detection horizon increases and a higher number of gravitational-wave (GW) events are positively identified as binary mergers. %
The events observed so far lie at distances $z\lesssim 1$~\cite{GWTC-3}, while the vast majority of binary mergers is expected to lie well beyond this horizon, as suggested by both theoretical expectations for the merger rate redshift evolution~\cite{Dominik:2014yma, Belczynski:2016obo,Rodriguez:2017pec} and its inferred trend from GW data through the Third Grvitaitonal-Wave Transient Catalog (GWTC-3)~\cite{pops-GWTC3}. %
This knowledge has motivated studies and inference of the sub-threshold, unresolved collection of binaries, treated as an overall gravitational-wave background (GWB) signal. %

Works forecasting the GWB from compact binaries populate the literature: before the first GW detections, these were based on theoretical models of the binary population~\cite{Phinney:2001di, Regimbau:2007cw, Regimbau:2011rp}, while more recently GW data--informed projections~\cite{iso-O3} have become a benchmark for the GW community. An important distinction to make is between estimates of a specific realisation of the background, for example used in mock-data challenges~\cite{Meacher:2014aca}, and estimates of the {\it ensemble average} of the background, which corresponds to the expectation value of the background amplitude targeted by GWB searches~\cite{Zhu_2011,PhysRevD.85.104024}. %
In both cases, the calculation involves the (expected) distributions for the individual binary parameters, such as mass, distance, and event rate distributions. %
These population models are often taken to be simple parametric functions with a well-defined set of hyper-parameters, which are fixed to fiducial (or assumed) values for the background calculation. Re-calculating the background signal for varying population hyper-parameters can become computationally intensive, when employing large sample sets.

Several applications in the literature require marginalizing the background signal over possible population configurations, including forecasting studies, such as those presented by the LIGO-Virgo-KAGRA (LVK) collaboration in~\cite{iso-O3,pops-GWTC3}, and inference analyses, such as~\cite{Callister:2020arv,iso-O3,Turbang:2023tjk}. %
As the interest in this type of work grows, there is a need for efficient and flexible background estimation procedures. %
In this paper, we present a method to efficiently carry out these calculations, and make an open-access code base, {\tt popstock}, available to the community. %
The novelty of our approach is in the design of a re-weighting technique which allows us to sample the binary parameter probability distributions and evaluate a corresponding set of waveform approximants only once, enabling an extremely efficient re-estimation of the GWB when varying population hyper-parameters. %
An analogous re-weighting approach was previously implemented in~\cite{Turbang:2023tjk}. %
Furthermore, we implement the use of waveform templates imported from the LIGO Scientific Collaboration Algorithm Library (LAL)~\cite{2020ascl.soft12021L}. 
To improve efficiency, previous codes have employed analytic waveform approximations directly embedded in the codebase; with our CPU--optimized spectral calculations and our GPU--optimized re-weighting technique, we are able to support a much broader range of waveforms through the commonly used python library for GW analysis, Bilby~\cite{Ashton:2018jfp}. 

We use {\tt popstock} to investigate the impact of the uncertainty on the binary population on the detection prospects of the GWB with ground-based interferometers. %
In particular, we compare with and extend work done in~\cite{pops-GWTC3} to include more uncertainties on the redshift distribution of sources. %
We also employ the package to probe the effect of waveform choice on the estimation, and whether this is entirely degenerate with the population uncertainty. %
We find that the expected background amplitude can be significantly boosted, when admitting higher mass mergers and higher rates of mergers at low-redshift, within current population uncertainties. %
We further find that the choice of waveform can have noteworthy effects on background signal estimates, however these do not dominate current population uncertainties.

This paper is organized as follows: In Sec.~\ref{sec:GWB} we introduce the theoretical aspects behind compact binary GWB calculations; in Sec.~\ref{sec:models} we define the analytic models used here to describe the distributions of black hole masses and distances; in Sec.~\ref{sec:popstock} we introduce {\tt popstock}, our new package for GWB calculations, outlining its key functionalities; in Sec.~\ref{sec:results} we present our background calculations and investigations; and finally we summarise our conclusions in Sec.~\ref{sec:conclusions}.


\section{The compact binary GW background}\label{sec:GWB}


The amplitude of the GWB signal is parametrized by the fractional energy density spectrum emitted by GWs throughout the Universe, $\OmegaGW(f)$, \cite{Phinney:2001di}%
\begin{equation}
	\label{eq:Omega}
	\OmegaGW(f) = \frac{1}{\rho_c}\frac{d\rho_\textsc{gw}}{d\ln f}\,,
\end{equation}
which is normalized by the critical energy density of the Universe $\rho_c=3c^2 H_0^2/8\pi G$. %
This is in general the {\it total} energy density contributed by GWs throughout the Universe, and is not restricted to sub-threshold signals. %
Calculating a residual background requires the definition of a detection cutoff, which is detector--dependent\footnote{For discussions on residual backgrounds, in particular in the context of next generation ground-based interferometers, we refer the reader to~\cite{Sachdev:2020bkk,Zhou:2022nmt, Bellie:2023jlq}, for example.}, while here and elsewhere $\OmegaGW$ is considered to be an astrophysical property of the Universe. %

While the GWB spectrum is detector--independent, it is useful to employ the detector frame to perform calculations. %
This allows us to immediately relate the intrinsic signal amplitude with the measured signal in GW detectors. %
As shown for example in~\cite{Meacher:2015iua, sym14020270}, the energy density spectrum $\OmegaGW(f)$ may be estimated as the average fractional GW energy density present in a detector during an observation time $\Tobs$,
\begin{equation}
    \OmegaGW(f) = \frac{f^3}{\Tobs} \frac{4\pi^2}{3H_0^2} \sum_i^{N_i} {\cal P}_d (\Theta_i; f)\,,
\end{equation}
assuming a finite number of GWs received at the detector, $N_i$. %
Here, ${\cal P}_d(\Theta_i, f)$ is the Fourier domain unpolarized power in the detector frame (hence the subscript $d$) associated to a GW with parameters $\Theta_i$, defined as
\begin{equation}\label{eq:Pdtheta}
    {\cal P}_d(\Theta_i; f) = \tilde{h}_{+}^2(\Theta_i; f) + \tilde{h}_{\times, i}^2(\Theta_i; f)\,,
\end{equation}
where $\tilde{h}_{A}(\Theta_i; f)$ is the Fourier transform of the GW waveform evaluated at $\Theta_i$\footnote{In the case of a stochastic signal described as a superposition of plane waves, as is the case often for cosmological signals, the signal can be thought of as a wave where the Fourier amplitudes are stochastic fields, the parameters $\Theta$ describe the field, and the power is the second-order moment of the field~\cite{Renzini:2022alw}.}. %
So defined, ${\cal P}_d$ has units $s^{-2}$.

In the limit of infinite observation time and infinite events, the $\OmegaGW$ spectrum approaches its ensemble average, $\barOmegaGW$. %
While the $\OmegaGW$ measured by an experiment depends on the specific realisation observed throughout the experiment observation time, $\barOmegaGW$ only depends on the distributions which describe the GW parameters $\Theta$, 
which are in turn parametrized via population hyper-parameters $\Lambda$~\cite{Phinney:2001di}. %
The $\barOmegaGW$ spectrum is thus a unique population signature, and is targeted in observations in practice by measuring $\OmegaGW$ for large $\Tobs$. %
The equivalence is easily seen in the limit of large GW numbers as
\begin{equation}
    \frac{1}{\Tobs}\sum_i^{N_i} {\cal P}_d (\Theta_i; f) \underset{\substack{N_i\rightarrow \infty\\
    \Tobs\rightarrow \infty}}{\approx} \frac{dN}{dt} \int d\Theta \, p_d(\Theta|\Lambda) \, {\cal P}_d (\Theta; f)\,, 
\end{equation}
where $p_d(\Theta|\Lambda)$ are the (normalized) detector frame probability distributions for the GW parameters $\Theta$, such that
\begin{equation}\label{eq:Omegabar}
    \barOmegaGW(\Lambda; f) = f^3 \frac{4\pi^2}{3H_0^2} R \int d\Theta \, p_d(\Theta|\Lambda) \, {\cal P}_d (\Theta; f)\,.
\end{equation}
Here, $R\equiv\frac{dN}{dt}$ is the total rate of events per unit detector-frame time. %
It can be convenient to isolate the redshift integral in Eq.~\eqref{eq:Omegabar}, assuming redshift is independent from other parameters, and incorporate the rate in the redshift evolution probability $p(z)$, defining the event rate per unit detector-frame time per redshift shell,
\begin{equation}
    R(z) = R\,p(z)\,. 
\end{equation}


The $\barOmegaGW$ spectrum can then be interpreted as an integral over redshift shells of the average GW power present in each shell, analogously to Eqs. (4) and (5) of~\cite{Callister:2020arv},
\begin{equation}\label{eq:Omegabar_rate}
    \barOmegaGW(\Lambda; f) = f^3 \frac{4\pi^2}{3H_0^2} \int dz\, R(z|\Lambda_z) \langle \, {\cal P}_d (z, \theta; f) \rangle\,,
\end{equation}
where the $\langle . \rangle$ brackets imply the GW spectral power samples ${\cal P}_d$ are averaged over the ensemble described by the parameter probability distributions, in each redshift shell. To see how the GW spectral power is related to the energy spectrum, for each binary, see App.~\ref{app:dEdf} for a pedagogical derivation. 


\section{BBH population models}\label{sec:models}

We illustrate the effect of the population model on $\Omega_{\rm GW}$ by adopting two phenomenological mass distribution models used in~\cite{pops-GWTC2, pops-GWTC3}. The \textsc{Powerlaw+Peak} model (PLPP), first introduced in \cite{Talbot18}, has been widely adopted in the literature as an astrophysically-motivated mass distribution model. The PLPP model consists simply of a truncated power-law, motivated by the shape of the stellar initial mass function, and a gaussian bump (or peak), originally intended to account for a possible overdensity of black holes around a certain mass, as motivated by, e.g., pulsational pair instability effects \citep{Talbot18, pops-GWTC2, pops-GWTC3} \footnote{This feature is found in the data, but recent works have cast doubt on whether it can be attributed to the pulsational pair instability mechanism \citep{Golomb23, Hendricks23}}. We list the parameters of the PLPP model in Table~\ref{tab:plpp}. In addition to PLPP, we also consider a simpler mass distribution, consisting of a truncated power-law with a break at a particular mass. While~\cite{pops-GWTC3} finds that this broken power-law (BPL) model is disfavored with respect to the PLPP model, we also consider it for illustration purposes. Parameters used for the BPL model are described in Table~\ref{tab:bpl}.

\begin{table}[h]
\centering
\begin{tabular}{c|p{6.5cm}}
\hline
Parameter & Description \\
\hline
$\alpha$ & Slope of the primary mass power-law. \\
\hline
$\beta$ & Slope of the mass ratio power-law. \\
\hline
$m_{\text{min}}$ & Minimum mass allowed in the system. \\
\hline
$m_{\text{max}}$ & Maximum mass allowed in the system. \\
\hline
$m_{\text{pp}}$ & Location of the Gaussian bump in the mass distribution. \\
\hline
$\sigma_{\text{pp}}$ & Width of the Gaussian bump in the mass distribution. \\
\hline
$\lambda$ & Fraction of sources in the bump. \\
\hline
\end{tabular}
\caption{Description of power-law-plus-peak (PLPP) model parameters.}
\label{tab:plpp}
\end{table}

\begin{table}[h]
\centering
\begin{tabular}{c|p{6.5cm}}
\hline
Parameter & Description \\
\hline
$\alpha_1$ & Slope of the primary mass power-law before the break. \\
\hline
$\alpha_2$ & Slope of the primary mass power-law after the break. \\
\hline
$m_{\text{min}}$ & Minimum mass allowed in the system. \\
\hline
$m_{\text{max}}$ & Maximum mass allowed in the system. \\
\hline
$m_{\text{pp}}$ & Location of the Gaussian bump in the mass distribution. \\
\hline
$b$ & Fractional distance between $m_{\rm min}$ and $m_{\rm max}$ of the break. \\
\hline
\end{tabular}
\caption{Description of broken-power-law (BPL) model parameters.}
\label{tab:bpl}
\end{table}

We also adopt redshift distribution models commonly used in the literature. For example, \cite{Fishbach:2018edt} introduced a broken power-law model to describe the merger rate as a function of redshift. This model is motivated by the observed star formation rate (SFR) across redshift: a rate rising to and peaking at some redshift ($z_{\rm peak}$, \citep{Mandel2016, Fishbach:2018edt}) then decaying down to high redshifts, where the star formation rate was much lower. A SFR model commonly adopted in the literature is the broken power-law fit from \cite{Madau:10.1146}, which is parameterized in terms of a low-redshift power-law index $\gamma$, high-redshift (negative) index $\kappa$, and a peak or turn-over redshift parameter $z_{\rm peak}$:
\begin{equation}
    {\cal R}_{\rm MD}(z) \propto \frac{(1+z)^\gamma}{1+\left(\frac{1+z}{1+z_{\rm peak}}\right)^\kappa}
\end{equation}
In the following, as in~\cite{pops-GWTC3} and~\cite{iso-O3}, we use ${\cal R}_{\rm MD}$ to describe the number of mergers per unit comoving volume $V_c$ per unit source-frame time $t_s$. %
For brevity, we refer to this model as MD. %
The parameters employed throughout for the MD model and their default values are shown in Table~\ref{tab:powerlawredshift}. %
In reality, the merger rate of BBHs from stellar collapse is a function of the SFR and the delay time between star formation and merger of the remnant~\cite{Fishbach:2021mhp, Callister:2020arv, Turbang:2023tjk}; this may make the number of mergers per unit comoving volume and source-frame time $R(z)$ deviate from a SFR-like broken power-law. 

\begin{table}[h]
\centering
\begin{tabular}{c|p{5.2cm}|c}
\hline
Parameter & Description & Defaults\\
\hline
$\gamma$ & Low-redshift power-law index. & 2.7 \\
\hline
$\kappa$ & High-redshift power-law index. & 5.6 \\
\hline
$z_{\rm peak}$ & Redshift of the peak rate. & 1.9 \\
\hline
\end{tabular}
\caption{Description of broken-power-law Redshift model parameters. Default values drawn from~\cite{Fishbach:2018edt, Callister:2020arv}. This model, including an overall local-merger-rate normalisation $R_0$, is referred to as MD throughout.}
\label{tab:powerlawredshift}
\end{table}

For illustration, we also adopt a less realistic model in which the merger rate is constant across cosmic time, ${\cal R}(z)=R_0$. This Uniform in Comoving Voume (UICM) model assumes that the merger rate as measured in the source frame of the emitter, is constant across redshift. %

The redshift of individual CBC merger events in the detector frame is drawn from a probability distribution $p(z)$ which takes into account the comoving volume per unit redshift gradient, $dV_c/dz$, and the redshifting of the rate from source-frame to detector-frame, 
\begin{equation}
    p(z) = \frac{1}{1+z} \frac{dV_c}{dz} {\cal R}(z)\,.
\end{equation}
The $p(z)$ functions for the MD and UICM rate evolution are illustrated in the left panel of Fig.~\ref{fig:redshift_models}.


In this paper, unless otherwise specified we draw uncertainties from the LVK collaboration GWTC-3 population posteriors, published in the data release~\cite{gwtc3_zenodo} 
which accompanied the collaboration results~\cite{pops-GWTC3}. The release includes samples from the posterior of population hyperparameters inferred through GWTC-3 (i.e., the population parameters governing the shapes of the mass, spin, and redshift distributions). We use these hyperparameter samples for the corresponding redshift and mass models described above in the analysis that follows.
We use the PLPP mass model and MD redshift model as fiducial models for our studies. The PLPP model is considered a good parametric description of the mass spectrum of GWTC-3~\cite{pops-GWTC3}, also confirmed by non-parametric approaches~\cite{PhysRevX.14.021005}, and has been widely used since in the context of GWB estimation (e.g., ~\cite{Zhou:2022nmt, Bellie:2023jlq, Sah:2023bgr}), while the MD model is a well-motivated astrophysical model~\cite{Fishbach:2018edt} and has been already employed in stochastic inference analyses~\cite{iso-O3}.


\section{{\tt popstock}}\label{sec:popstock}
We present {\tt popstock}, a python--based open--source package for the rapid computation of background spectra such as $\OmegaGW$, for a given realization of events, and $\bar{\Omega}_{\rm GW}$, for a given set of hyper-parameters $\Lambda$. %
Other than the standard python scientific libraries {\tt numpy}~\cite{numpy} and {\tt scipy}~\cite{scipy}, the main dependencies of the {\tt popstock} package are: {\tt astropy}~\cite{astropy}, a core python library used by astronomers; {\tt bilby}~\cite{Ashton:2018jfp}, a most popular bayesian inference library for GW astronmy; and {\tt gwpopulation}~\cite{gwpopulation}, a hierarchical bayesian inference package containing a collection of parametric binary black hole mass, redshift, and spin population models. %

The {\tt popstock} package relies on {\tt multiprocessing} (included in most python distributions) to parallelize the computation of $\OmegaGW$ for large $N_i$. %
The GW waveforms required to compute ${\cal P}_d(\Theta_i; f)$ in Eq.~\eqref{eq:Pdtheta} are evaluated at each $\Theta_i$ using the {\tt bilby} library, which in turn imports LAL~\cite{2020ascl.soft12021L}. %
This allows us to employ a vast array of modern waveforms in our computations. %

To compute $\bar{\Omega}_{\rm GW}$ for a given set of population hyper-parameters $\Lambda$ and a given collection of population models, we directly sample the probability distributions $p_d(\Theta|\Lambda)$ and evaluate Eq.~\eqref{eq:Omegabar} via a Monte Carlo simulation. %
The accuracy of this evaluation depends on the number of samples employed, as discussed below. %
This approach is limited by the long sampling and evaluation times of the GW waveforms, and is not an optimal tool to perform in-depth studies of the impact of population uncertainties on the $\bar{\Omega}_{\rm GW}$ spectrum. %
Hence, {\tt popstock} includes a re-weighting technique to compute $\bar{\Omega}_{\rm GW}$ for a new set of $\Lambda$ parameters without re-evaluating Eq.~\eqref{eq:Omegabar}. %

In the rest of this section, we describe the {\tt popstock} re-weighting technique and probe its efficiency and accuracy.

\begin{figure*}
    \centering
    \includegraphics[width=\textwidth]{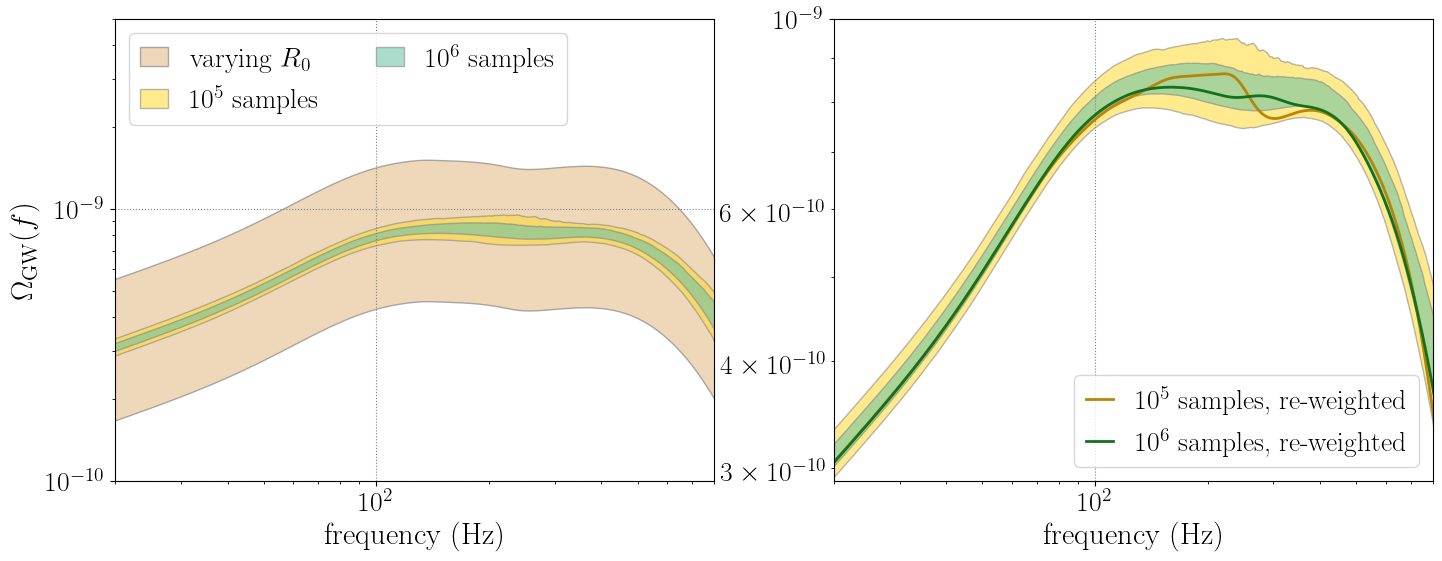}
    \caption{Impact of sample variance and re-weighting on the $\OmegaGW$ spectrum. On the left: 95\% confidence on the spectrum calculated using $10^5$ ($10^6$) samples in yellow (green) drawn from a fixed hyper-parameter distribution, compared to the 95\% confidence on the spectrum including uncertainty on the local merger rate parameter in brown. Unsurprisingly, the sample variance varies greatly with number of samples and when using large sample numbers is subdominant compared to population parameter uncertainty. On the right: sample variance from the left panel compared to re-weighted estimates of the $\OmegaGW$ spectrum. The re-weighted spectra lie neatly within the sample variance uncertainty bounds, implying that a re-weighted spectrum is undistinguishable from a regularly sampled spectrum, with these sample numbers.}
    \label{fig:sample_variance}
\end{figure*}
\subsection{Re-weighting methodology}
We lay out a simple method to efficiently calculate $\barOmegaGW$ for different sets of hyper-parameters $\Lambda_i$ describing the (detector frame) target population distributions, $p_{d}(\Theta|\Lambda_i)$. %
The integral of~\eqref{eq:Omegabar} above allows the implementation of an importance sampling approach or {\it re-weighting}\footnote{Reweighting has become a popular tool for efficient Monte Carlo computations in GW astronomy. See, e.g., \cite{gwpopulation, Payne2019, Hourihane23, Talbot23} for a review of some appplictions of this method in the GW field.}, whereby
\begin{equation}\label{eq:reweight}
    \begin{aligned}
         \int d\Theta \, p_{d}(\Theta|\Lambda_1) \, {\cal P}_d (\Theta) &= \int d\Theta \, \frac{p_{d}(\Theta|\Lambda_1)}{p_{d}(\Theta|\Lambda_0)} p_{d}(\Theta|\Lambda_0) \, {\cal P}_d (\Theta)\\ &\equiv \int d\Theta \, w_{0 \rightarrow 1}(\Theta) p_{d}(\Theta|\Lambda_0) \, {\cal P}_d (\Theta)\,,
    \end{aligned}
\end{equation}
where $p_{d}(\Theta|\Lambda_0)$ is a chosen as the fiducial distribution, and $w_{0 \rightarrow 1}$ is the {\it weight} required to ``transform'' between the fiducial distribution and the target one, relative to $\Lambda_1$:
\begin{equation}
    w_{0 \rightarrow 1} =\frac{p_{d}(\Theta|\Lambda_1)}{p_{d}(\Theta|\Lambda_0)}\,.
\end{equation}
In practice, this reweighting approach is more efficient than direct Monte Carlo integration when $p_d(\Theta|\Lambda_1)$ is hard to sample from but easy to evaluate. Therefore, we first draw a large set of samples $\bold{\Theta}$ from the fiducial population $\Lambda_0$ and compute the probability of drawing those samples $p_d(\Theta|\Lambda_0)$. This is stored as the denominator in the weights $w$. Each time the integral of Eq.~\ref{eq:reweight} is evaluated for some \textit{different} population $\Lambda_1$, it is only necessary to evaluate the probability of those fiducial samples under the target population $p_d(\Theta|\Lambda_1)$, as $w$ is the only term that depends on $\Lambda_1$. %
See also App.~D of~\cite{Turbang:2023tjk} for an analogous re-weighting approach to calculate the background spectrum.

The re-weighting operation is directly implemented in the Monte Carlo evaluation described above, which is valid as long as a sufficient number of samples $\Theta_j$ are used,  
\begin{equation}
    \int d\Theta \, w_{0 \rightarrow 1}(\Theta) p_{d}(\Theta|\Lambda_0) \, {\cal P}_d (\Theta) \approx \sum_j   w_{0 \rightarrow 1} (\Theta_j)  {\cal P}_d (\Theta_j)\,. 
\end{equation}
This allows us to evaluate the (costly) ${\cal P}_d$ spectra {\it only once}, and re-weight the contribution of each wave according to a desired target distribution.


In practice, we rely on the source-frame population distributions to sample the GW parameters. To convert these to detector frame, we evaluate the relevant jacobians assuming a fixed cosmology,
\begin{equation}
    p_d(\Theta|\Lambda) = p_s(\Theta|\Lambda) \,\frac{d\Theta_s}{d\Theta_d}\,.
\end{equation}

\subsection{Effective sample size and sample variance}\label{sec:samples}


As a check of the performance of our re-weighting approach, we estimate the effective sample size $N_{\rm eff}$ for different number of samples $N_{\rm s}$ and different $\Omega_{\rm GW}$ spectra, and ensure $N_{\rm eff}\gg 1$. %
We find that, for a fixed re-weighting set of $\Lambda$ hyper-parameters, $N_{\rm eff}\approx2\times 10^4$ for $N_{\rm s}=5\times 10^4$, $N_{\rm eff}\approx4\times 10^4$ for $N_{\rm s}=5\times 10^4$, and
$N_{\rm eff}\approx3\times10^5$ for $N_{\rm s}= 10^6$. %
In practice, these numbers will depend on the size of the parameter space probed by the re-weighting. %
In all implementations shown in this paper we use $N_{\rm s}=1\times 10^6$ unless otherwise stated, and we have checked the order of magnitude of $N_{\rm eff}$ reported here remains reliable for all results shown.

The $\Omega_{\rm GW}$ spectrum is by definition a stochastic observable, and thus presents an intrinsic sample variance. %
In particular, $\Omega^{\rm BBH}_{\rm GW}$ is dominated by a  Poisson process given the low merger rate of black hole binaries. %
We estimate the intrinsic variance of $\Omega^{\rm BBH}_{\rm GW}$ associated to different number of samples (which can be directly converted to different observation times, assuming a total merger rate) by calculating the spectrum using different sample draws from a fixed set of population priors. %
These are shown in Fig.~\ref{fig:sample_variance} (left panel) using $10^5$ and $10^6$ samples, where the shading indicates the 95\% interval over 1000 sets. %
In this case, increasing the number of samples by a factor of 10 decreases the width of the 95\% interval by 52\% on average, for frequencies between $20-500$ Hz. %
This specific example corresponds to the following set of hyper-parameters for the PLPP mass model: $\alpha=3.5$, $\beta=1$, $\delta_m=4.5$, $\lambda=0.04$, $m_{\rm max}= 100$, $m_{\rm min}= 4$, $m_{\rm pp}=34$, $\sigma_{\rm pp}=4$. The redshift model is fixed to the default MD model defined above with a local merger rate of $R_0=15$ Gpc$^{-3}$yr$^{-1}$. %

We compare this intrinsic uncertainty to re-weighting: as may be seen in Fig.~\ref{fig:sample_variance} (right panel), re-weighted curves for $\OmegaGW^{\rm BBH}$ for the given set of $\Lambda$ hyper-parameters lie within the 95\% sample uncertainty on the spectrum, for different values of $N_{\rm s}$. %
This implies the re-weighted $\OmegaGW^{\rm BBH}$ spectrum for a given population model is within the intrinsic error on that spectrum, and is thus a fair approximation to make.

As we focus on BBHs in this paper, we drop the BBH subscript from $\OmegaGW^{\rm BBH}$ in what follows, and assume we refer to the BBH spectrum unless otherwise specified.


\section{Background projections using {\tt popstock}}\label{sec:results}

We study the dependence of the amplitude, spectral shape, and uncertainty of $\OmegaGW$ on various models and data products. %
These studies will fundamentally inform compact binary population parameter estimation campaigns with upcoming GW datasets, for example in the style of~\cite{Callister:2020arv, iso-O3}, which use constraints on (or, in the future, measurements of) the $\OmegaGW$ spectrum.

We consider here a frequency range of $10-2000$ Hz as this corresponds to the sensitivity of second generation ground-based GW detectors such as the current configurations of the LIGO, Virgo, and KAGRA instruments as well as their near-future improvements.

\subsection{Mass and redshift models}\label{sec:mass_redshift}
\begin{figure*}
    \centering
    \includegraphics[width=\textwidth]{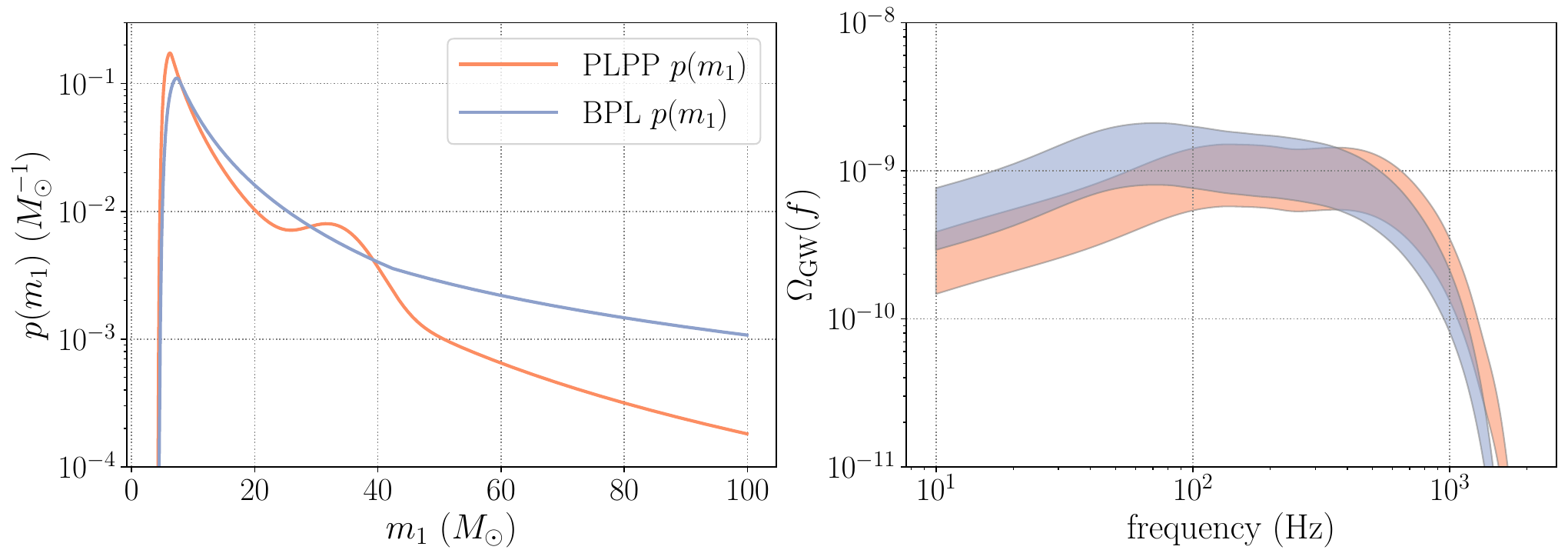}
    \caption{Impact of the primary mass distribution on the $\OmegaGW$ spectrum. On the left, the two primary mass model probability densities used throughout; on the right, 95\% confidence intervals for $\OmegaGW$ using the two mass models, including uncertainty on the local merger rate from~\cite{pops-GWTC3}. }
    \label{fig:mass_models}
\end{figure*}
\begin{figure*}
    \centering
\includegraphics[width=\textwidth]{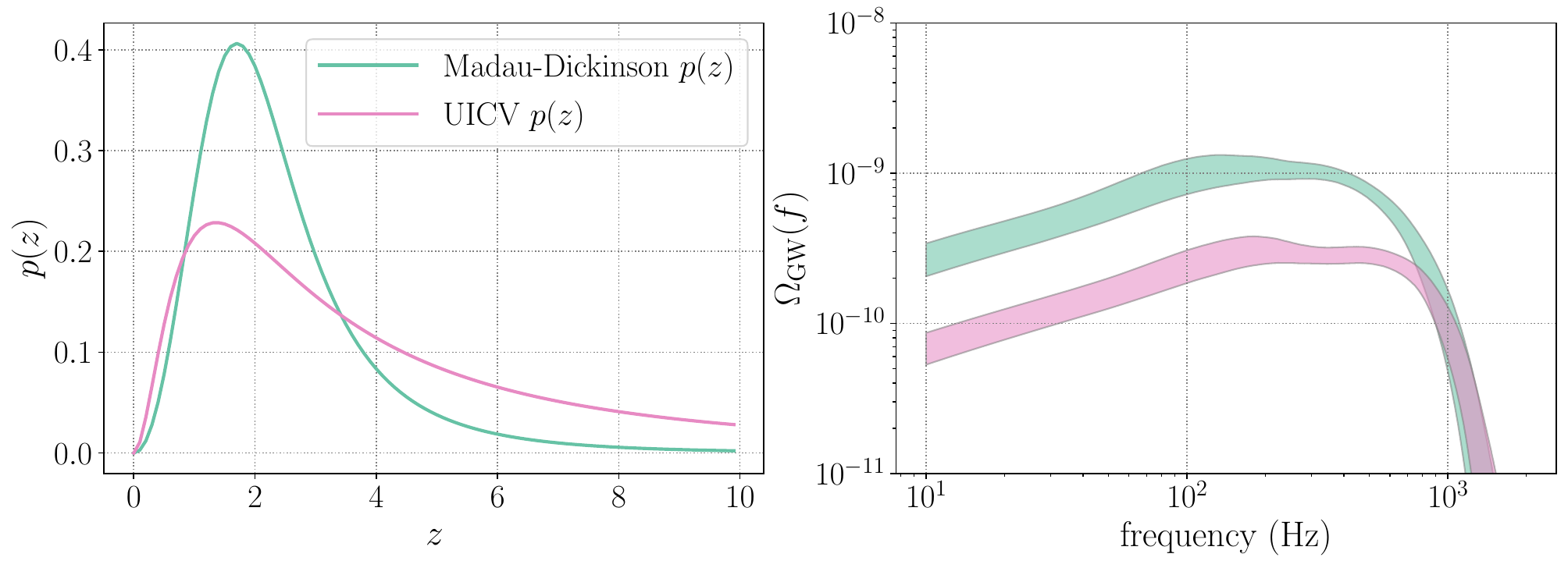}
    \caption{Impact of the merger rate redshift distribution model on the $\OmegaGW$ spectrum. On the left, the two redshift evolution model probability densities used throughout; on the right, 95\% confidence intervals for $\OmegaGW$ using the two merger rate models, including uncertainty on the PLPP mass model from~\cite{pops-GWTC3} as described in Sec.~\ref{sec:models}. }
    \label{fig:redshift_models}
\end{figure*}
With {\tt popstock} we can rapidly assess the impact of different mass and redshift models on the projected $\OmegaGW$. %
Here we compare the PLPP and BPL mass models, and the UICV and MD redshift models, all  introduced in Sec.~\ref{sec:models}. %
When comparing mass models, we fix the mass model hyper-parameters while including the uncertainty on the local merger rate $R_0$ from~\cite{pops-GWTC3}\footnote{Specifically: we employ samples from the posterior fit of the power-law redshift model, as in~\cite{pops-GWTC3} no broken-power-law redshift model was used.} assuming the MD redshift evolution with all other parameters fixed to the values discussed above. %
When comparing redshift models, we instead fix the redshift model hyper-parameters while including the uncertainty from~\cite{pops-GWTC3} on the PLPP mass model. %
 We deliberately choose values for certain $\Lambda$ hyper-parameters which are unrealistic and not favoured by current data to showcase the effect different mass and redshift models may have on the $\OmegaGW$ spectral shape. %

A comparison between the PLPP and BPL mass models is shown in Fig.~\ref{fig:mass_models}. %
The PLPP parameters are fixed to the same set used in Sec.~\ref{sec:samples}, and BPL to $\alpha_1=-2$, $\alpha_2=-1.4$, $\beta=1$, $b=0.4$ (see Sec.~\ref{sec:models} for details on the parameters). %
As PLPP is commonly used as a mass model when generating $\OmegaGW$ forecasts (as in~\cite{iso-O3, pops-GWTC3}), we take this as the fiducial model to produce $\OmegaGW$ spectra and compare those obtained with the BPL mass model against these. Note that in particular the choice of $\alpha_2>\alpha_1$ for BPL here implies a larger amount of high-mass binaries in the distribution, as seen in the left panel of Fig.~\ref{fig:mass_models}, where primary mass probability distributions are shown. %
These more massive binaries merge at lower frequencies and their emission is further redshifted into the lower end of the frequency range considered here; remember that, for example, we detect an equal-mass binary with true component masses of 70$M_\odot$ merging around $\sim 60$ Hz at $z=0$ and $\sim 30$ Hz at $z=1$ (see e.g.~\cite{Renzini:2024hiu} for more considerations along this line). 
As seen in the right panel of Fig.~\ref{fig:mass_models}, this both boosts the amplitude of $\OmegaGW$ at all frequencies below a few hundred Hz, and changes the spectral shape at these frequencies, when compared to the PLPP mass spectrum. %
Specifically, the typical ``turnover'' in the spectrum corresponding to the frequency at which most binaries have merged is  broken into two turnovers: one for the higher mass binaries (below 100 Hz) and one for the lower mass ones (around 300 Hz). %
This is effect is certainly fuelled by the unrealistic parameter choice made for BPL ($\alpha_2>\alpha_1$). %
Comparatively, the PLPP model gives rise to a single turnover with a plateau between $\sim 100=400$ Hz, which presents small features (``wiggles'') which are related to the redshifting of the peak at $33M_\odot$. %
Futhermore, the spectral index at lower frequencies is more peaked than that of the PLPP model. %
A simple broken-power-law fit to the two sets of curves shown in the right panel of Fig.~\ref{fig:mass_models} yields $\alpha=0.61$ and $\alpha=0.76$ for the lower frequency region of the $\OmegaGW$ spectrum corresponding to the PLPP and BPL mass models, respectively. %
A broader discussion on data-informed spectral indices is postponed to Sec.~\ref{sec:O3_results}. %

The uncertainty on the local rate $R_0$ implies that there is significant overlap between the 95\% credible envelope of the spectrum from these two mass models, suggesting it would be challenging to distinguish mass spectrum features from redshift ones from a measurement of $\OmegaGW$ alone. %
The overlap would be even greater when including full uncertainty on the redshift model parameters. %
However, if a large amplitude and large spectral index (i.e., $\alpha>2/3$) $\OmegaGW$ is observed at low frequencies, we expect a mass model which admits large mass binaries (such as the BPL one showed here) to be favoured.

In Fig.~\ref{fig:redshift_models} we compare the effect of the UICV and MD redshift models on $\OmegaGW$. %
We fix the local merger rate to $R_0=15$ Gpc$^{-3}$yr$^{-1}$, and compare a UICV rate evolution to the default MD evolution (see Sec.~\ref{sec:models}) while we include the uncertainty on the PLPP mass model from~\cite{pops-GWTC3}. %
Most notably, the UICV model impacts the overall amplitude of the $\OmegaGW$ spectrum across all frequencies. %
In this test case, the decrease in amplitude when assuming UICV is approximately constant (and equal to a factor of $\sim 4$) between $10-100$ Hz, and is due to the lower merger rate between $1<z<4$. %
This effect is much larger than the impact of the uncertainty on the mass model, confirming that a measurement of $\OmegaGW$ will have significant information on the merger rate redshift evolution (as also seen in~\cite{Callister:2016ewt,Callister:2020arv, Renzini:2024hiu}). %
The turn-over in the spectrum appears shifted to slightly higher frequencies in UICV, possibly due to the slightly higher rate fraction at low redshift compared to MD, which instead increases as a power-law\footnote{This is evident in the area on the left panel of Fig.~\ref{fig:redshift_models} where the UICV model $p(z)$ lies above the MD $p(z)$, at $z<1$.}. %
Otherwise, the different redshift models do not appear to cause large variations in the overall shape of the spectrum, suggesting that the mass spectrum dominates these features. %

A natural extension of this study is to consider BBH mass spectra that evolve with redshift: this will mix the effects seen here when considering independent contributions, and in principle will need to be appropriately included in parameter estimation studies to avoid biases. %


\subsection{Waveform approximants}\label{sec:WFs}
The choice of waveform approximant, while central in certain studies of individual compact binary merger events, has been explored very little in the context of the compact binary background signal. %
In previous work (see e.g. approximations made in~\cite{Meacher:2015iua,Callister:2016ewt, Mukherjee:2021ags}), it was deemed sufficient to capture solely the evolution of the GW amplitude as a function of frequency, as the background $\OmegaGW$ contains no phase information; this evolution can be tracked analytically up to arbitrary Post-Newtonian (PN) order, considerably speeding up the calculation of $\OmegaGW$ compared to calculating full waveforms for large sets of events. Specifically, most works employ the amplitude component of frequency-domain inspiral-merger-ringdown (IMR) waveforms~\cite{Ajith:2007kx,Ajith:2009bn} defined analytically by parts, where the transition of the GW from one phase to the next is set by the specific intrinsic parameters of the binary (mass and spin). %
As the background has remained a weak signal in the current LVK data, a precise quantification of the systematic differences between background estimates with different waveform approximants has not been necessary. %
However, as detector sensitivities improve and detection becomes a real possibility, all modelling systematics are important to quantify (see also~\cite{Zhou:2022nmt}, ~\cite{PhysRevD.109.123014}). %
Here, we investigate  the effect of the waveform approximant using {\tt popstock} and confirm whether it is subdominant to the impact of population uncertainties.

\begin{figure}
    \centering
    \includegraphics[width=0.5\textwidth]{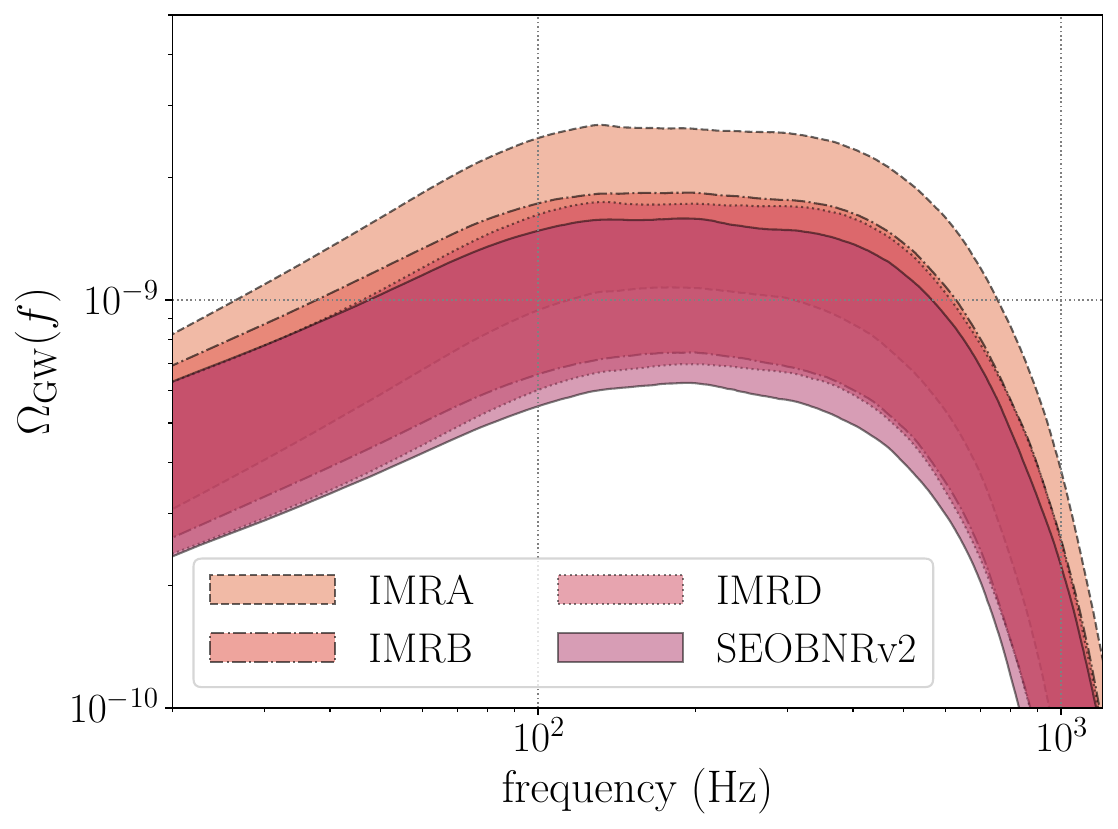}
    \caption{Impact of the waveform model on the $\OmegaGW$ spectrum. The shading indicates the 95\% confidence on the spectrum including the uncertainty on the PLPP mass model and the local merger rate, assuming a fixed MD redshift evolution.}
    \label{fig:waveforms}
\end{figure}

We focus here on the IMRPhenom family of waveforms commonly used in the literature to compute $\OmegaGW$, as well as an effective-one-body (EOB) numerical-relativity (NR) waveform model. In all cases, we omit spin effects, setting both black hole spins to 0. %
The specific waveforms used are
\begin{itemize}
    \item {\bf IMRPhenomA}~\cite{Ajith:2007kx}: The first IMR waveform, developed for GW data analysis in the frequency domain for non-spinning binaries. %
    Here the amplitude is expanded to leading post-newtonian (0-PN) order, implying the inspiral phase presents the characteristic $f^{2/3}$ trend (in $\OmegaGW$ units).
    \item {\bf IMRPhenomB}~\cite{Ajith:2009bn}: A direct successor of IMRPhenomA, this waveform includes higher order corrections in the amplitude term up to 1.5-PN and includes non-zero aligned spin. These correct the spectral shape of the waveform amplitude, as a function of both mass and spin.
    \item {\bf IMRPhenomD}~\cite{Khan:2015jqa}: A recent waveform including corrections up to 3-PN order in the amplitude and a more sophisticated fit to numerical relativity compared to IMRPhenomA and B.
    \item {\bf SEOBNRv2}~\cite{Purrer:2015tud}: An EOB NR waveform for spin-aligned BBHs, calculated numerically in the frequency domain.
\end{itemize}
A comparison between the $\OmegaGW$ spectra calculated for the same BBH population using the waveforms above is presented in Fig.~\ref{fig:waveforms}. %
We show the 95\% confidence contours on the population $\OmegaGW$ including the uncertainty on the mass model (PLPP) and the local merger rate $R_0$ assuming a fixed MD redshift evolution (as defined above). %

\begin{figure*}
    \centering
    \includegraphics[width=\textwidth]{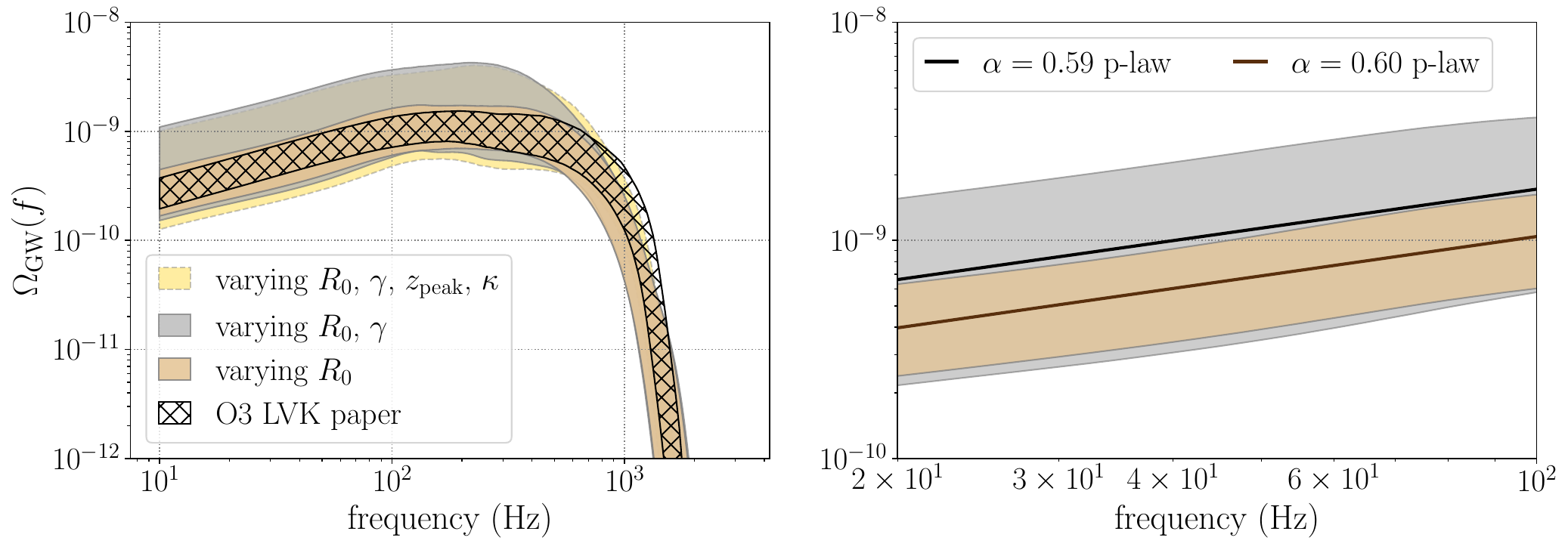}
    \caption{Uncertainty on the expected $\OmegaGW$ spectrum from BBHs due to uncertainty on the merger rate evolution parameters. On the left: 95\% confidence levels on the projected $\OmegaGW$ spectrum including uncertainty on the PLPP mass model, and assuming a MD merger rate model including progressively: uncertainty on the local merger rate $R_0$ (brown), uncertainty on the low-redshift spectral index $\gamma$ (gray), and finally uncertainty on the all redshift parameters (yellow). The hatched outline reports previous results published in~\cite{pops-GWTC3}. On the right: a zoomed-in comparison at low frequencies of the uncertainty on the $\OmegaGW$ spectrum when varying only $R_0$ and when varying $R_0$ and $\gamma$, reporting average spectral indices referred to these two contours. }
    \label{fig:omegas_O3}
\end{figure*}
We find that when employing the 0-PN IMRPhenomA waveform, the background signal is overestimated at all frequencies by up to 50\% in the range $10 < f <1000$ Hz compared to IMRPhenomB; this is due to the missing amplitude corrections to the inspiral phase (see e.g. the differences in Eq.~(4.13) of~\cite{Ajith:2007kx} and Eq.~(1) of~\cite{Ajith:2009bn}). While the amplitude estimates for IMRPhenomA and B agree at $f\equiv 0$, these diverge for $f>0$ as the amplitude evolution with frequency is $\Omega_A \propto f^{2/3}$ for IMRPhenomA and $\Omega_B \propto f^{\alpha < 2/3}$ for IMRPhenomB. The trend of $\alpha$ will depend on the specific mass, redshift, and spin realisation as discussed in Sec.~\ref{sec:mass_redshift}. %
This result shows that the somewhat basic assumption that the compact binary background at frequencies under $\sim 100$ Hz is well-approximated by a $2/3$ power-law can be upgraded, and informed by likely population models to optimize background searches.

Differences due to the use of IMRphenomB/D and SEOBNRv2 approximants are comparable to eachother and would be hard to distinguish from population uncertainties. %
Nevertheless, we comment that the different NR calibration used in IMRphenomD compared to B is evident in the impact due to the inspiral phase on the GWB signal, as the frequency evolution at low frequency is slightly modified, and SEOBNRv2 estimates an overall lower background than the IMRPhenom waveforms. 

The impact of including higher-order modes in the waveform calculation on the background spectrum was found to be negligible; a comparison between spectra calculated with the IMRPhenomD and IMRPhenomXPHM waveforms is included for completeness in App.~\ref{app:HOMs}.


\subsection{O3 population samples}\label{sec:O3_results}

We conclude our analyses by combining the uncertainties on the mass and redshift distributions drawn directly from the LVK GWTC-3 population analysis~\cite{pops-GWTC3}. %
We limit our focus to the MD redshift model for BBHs as this is the only model we have viable posterior samples for: in the case of the low-redshift merger rate parameters ($R_0$, $\gamma$), we use samples from the power-law redshift inference results released in~\cite{gwtc3_zenodo} for the power-law redshift model, while when including high-redshift features ($\zmax$, $\kappa$) we use results obtained performing inference on the entire MD model as done in~\cite{iso-O3}. %

Results obtained progressively varying the redshift hyper-parameters are shown in Fig.~\ref{fig:omegas_O3} (left panel). %
We include uncertainty over the entire PLPP mass hyper-parameter space, while varying: 
\begin{itemize}
    \item [(i)] just the local merger rate $R_0$, assuming a fixed broken-power-law merger rate evolution with parameters fixed to those discussed in Sec.~\ref{sec:models};
    \item[(ii)] both $R_0$ and the local power-law index $\gamma$, keeping the high-redshift parameters fixed;
    \item[(iii)] all parameters for the merger rate, including the turn-over redshift $z_{\rm peak}$ and high-redshift power-law index $\kappa$. 
\end{itemize}
In cases (i) and (ii) samples are drawn from the power-law redshift model posteriors of~\cite{pops-GWTC3}. %
In case (iii), samples are instead drawn from a full GWTC-3, O1--O2--O3 joint stochastic-population analysis (similar to~\cite{iso-O3}, for details on how this analysis is carried out see~\cite{Callister:2020arv}) as these also include posteriors on the higher redshift evolution of $R(z)$~\cite{callister_zenodo}. 

We compare the 95\% confidence levels on $\OmegaGW$ in case (i) with published results (shown in Fig.~\ref{fig:omegas_O3} in hatched outline\footnote{The hatched outline is exactly the green region highlighted in Fig.~23 of~\cite{pops-GWTC3}, which is publicly available in~\cite{gwtc3_zenodo}.}) and find these to be consistent. Note that the corresponding LVK contours draw from the same PLPP mass posterior and local merger rate posterior, 
but assume a different redshift evolution (see the original paper discussion~\cite{pops-GWTC3}), which explains the small differences between the curves at low frequency and the different turn-over trend at high frequency, which is dominated by low-redshift effects. %
The LVK contour was calculated assuming the IMRPhenomB analytic waveform model. %
Case (ii) and case (iii) give almost identical contours, which implies that the population analysis~\cite{pops-GWTC3} and the stochastic constraints~\cite{iso-O3} carry little information about the high-redshift evolution of the merger rate. %
Furthermore, these show that the uncertainty on the local merger rate evolution alone could account for an increase of up to a factor of $\sim 5$ in the $\OmegaGW$ spectrum. %
This could have considerable consequences on the detectability of the signal. %
In the right panel of Fig.~\ref{fig:omegas_O3} we zoom into the $[20,200]$ Hz portion of the (i) and (ii) spectra, and provide results of single power-law fits to the envelope of $\OmegaGW$ curves. %
The average $\alpha$ spectral indices found are consistent with eachother, $\alpha_{(i)}=0.59\pm0.02$ for (i), and $\alpha_{(ii)}=0.60\pm0.03$ for (ii). %
This confirms $\gamma$ has no impact on the low frequency spectral shape of $\OmegaGW$, but only on its amplitude.
\begin{figure}
    \centering
    \includegraphics[width=0.5\textwidth]{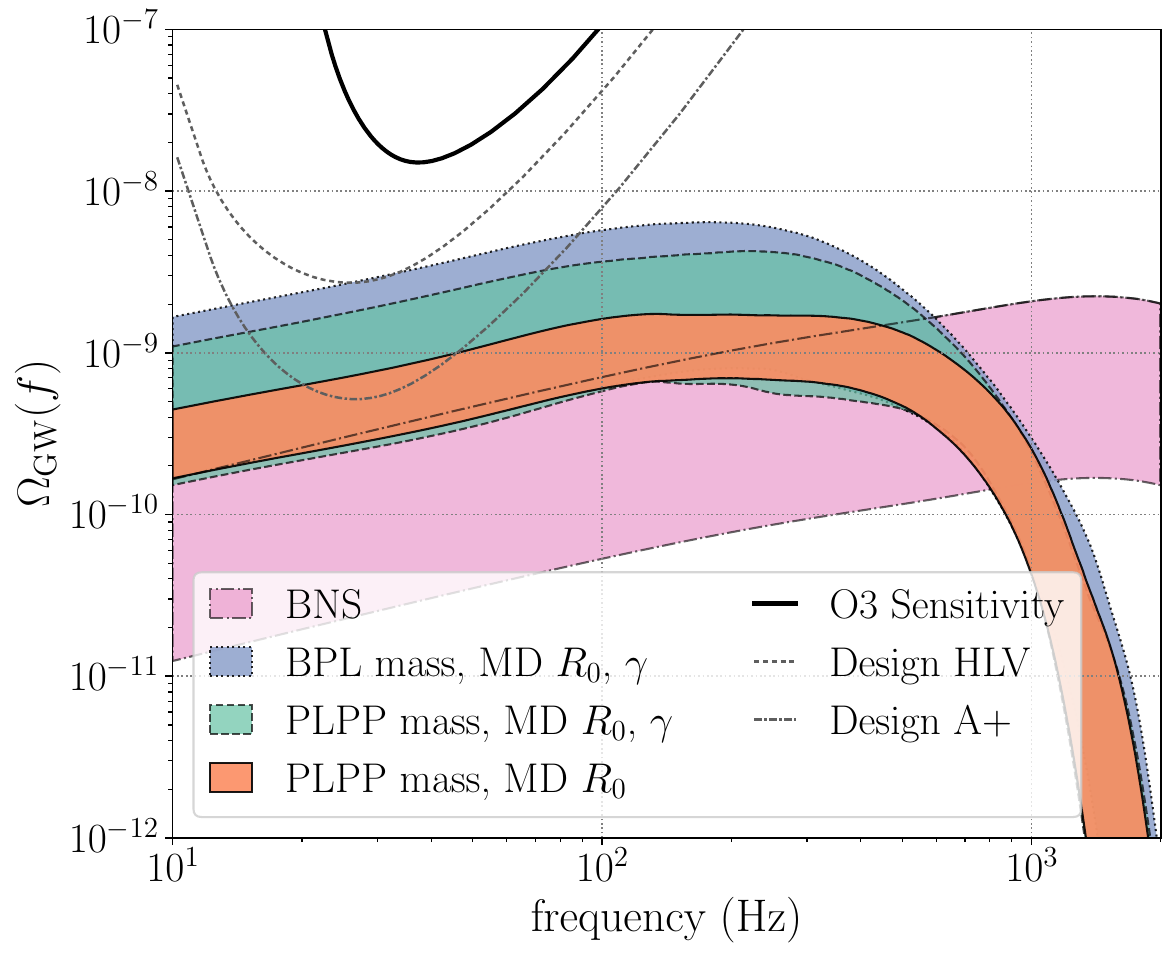}
    \caption{Projections of the background $\OmegaGW$ spectrum, given our knowledge on the compact binary population. Shaded regions outline 90\% credible bands for the GWB from BBHs and BNS (in pink), including uncertainty on the mass and redshift models for these sources using samples released in~\cite{pops-GWTC3, gwtc3_zenodo}. For BBHs we report uncertainty due to two mass models: the PLPP mass model, assuming a MD redshift model with uncertain local merger rate $R_0$ (orange), and also uncertain low-redshift power-law index $\gamma$ (green); and the BPL mass model, assuming a MD redshift model with uncertain $R_0$ and $\gamma$. Current and projected sensitivity
curves are included for reference. }
    \label{fig:detectability}
\end{figure}

We repeat the $\OmegaGW$ calculation varying $R_0$ and $\gamma$ using the BPL mass model, sampling over the joint mass and redshift posteriors obtained by performing population inference on the GWTC-3 catalog (samples are publicly available as a part of the example sample sets in the {\tt popstock} package repository\footnote{https://github.com/a-renzini/popstock}). We compare these results with the PLPP results described above in Fig.~\ref{fig:detectability}, overlaying the $2\sigma$ LVK power-law-integrated sensitivity curves~\cite{ThraneRomano2013} already shown in~\cite{pops-GWTC3}. %
These track the present and future upgrades to the LIGO and Virgo facilities, where ``O3 sensitivity'' is given by the O3 measured spectra, 
``Design HLV'' is produced assuming projections shown in~\cite{KAGRA:2013rdx} and is expected to approximate the sensitivity at the end of the O4 observing run (currently ongoing), and ``Design A+'' refers to the sensitivity projected for the next observing run O5 assuming 1 year of continuous data and all planned improvements to the detectors are implemented and successful~\cite{}. %

The BPL mass model predicts a systematically larger background, by a factor of 1.4 on average, which hints at the possibility of a louder signal than previously projected and thus the prospect of a detection of the stochastic signal before reaching the Design A+ LIGO--Virgo sensitivity. %
For completeness, we also include the expected background signal arising from binary neutron stars (BNS) in Fig.~\ref{fig:detectability}. %
This signal is strongly dominated by uncertainties given the very few detections of BNS mergers~\cite{BNS1, BNS2}. %
The projection is calculated employing the NRTidalv2 model discussed in~\cite{Dietrich:2019kaq}. %
We assume a uniform mass model between $1 - 2.5$ $M_\odot$, as in~\cite{pops-GWTC3}, and a merger rate model corresponding to a time-delayed SFR as used for projections presented in~\cite{iso-O3}\footnote{This model assumes the BNS progenitor formation rate is proportional to the SFR, and the distribution of time delays between binary formation and merger is inversely proportional to the time delays distributed between 20 Myr and 13.5 Gyr.}. We draw the local merger rate $R_0$ from the corresponding posterior samples presented in~\cite{pops-GWTC3,gwtc3_zenodo} -- for reference, we refer to the samples that set the $R_0=105.5^{+190.2}_{-83.9}$Gpc$^{-3}$yr$^{-1}$ upper limit.

\section{Conclusions}\label{sec:conclusions}

We present a novel method and code-base to rapidly calculate the background spectrum for inspiralling and coalescing compact binaries starting from a given population model and hyper-parameter sets. %

We quantify the joint uncertainty on the $\OmegaGW$ spectrum from both the mass and redshift distributions of the BBH population, given the most recent results from the LVK collaboration. %
Predictably, the uncertainty on the local merger rate and its evolution (together with the uncertainty on the mass model) dominate the expected amplitude of the spectrum, and can have significant implications on detectability.

Furthermore, we find that for the preferred mass model (power-law-plus-peak), the low-frequency spectral index of the stochastic background signal is $\alpha=0.6$. %
Previous detection approaches assumed $\alpha=2/3$; this result was based on the waveform used to calculate the expected GWB and its PN order expansion. %
We find that, when employing 0-PN order waveforms, there is a tension between projections for $\OmegaGW$ from the presently-observed population which competes with population uncertainty itself. %
The mismatch between the treatment of the late inspiral phase in IMR waveforms is particularly significant, as it is present throughout the entire $\OmegaGW$ spectrum and in particular at lower frequencies, where current detector sensitivities peak. %

Differences between specific background realisations and number of samples also produce different projections which may give rise to small tensions in the higher frequency range, where the spectra present a turn-over which is highly dependent on the binary mass distribution and local features of the merger rate. %
Current-generation detectors are not sensitive to this region of the spectrum, however this will have significant implications for next-generation interferometers.

In conclusion, the specific population properties of the CBC population as well as the specific realisation during our observations will play a role in detection capabilities. %
In particular, within current binary black hole population uncertainties, a low-redshift amplification of the merger rate and a larger population of higher-mass binaries contribute to a significant boost in the background amplitude, in the LVK sensitivity band, which could lead to early detection. %
On the other hand, more astrophysically-motivated BBH rate evolution models relate the merger rate to binary progenitor features, and re-parametrize the merger rate density in terms of, for example, the time-to-merger delay distribution and the host galaxy metallicity~\cite{timedelay_fishbach, metallicity_Chruslinska}. These models have recently been employed in joint analyses of the GWTC-3 catalog and LVK stochastic upper limits~\cite{Turbang:2023tjk}, and may provide alternative forecasts of the uncertainty on the $\OmegaGW$ spectrum, as we gather more GW data. %
These models will progressively be included and uptaded in {\tt popstock}. %
With {\tt popstock}, we provide the GW community engaged in GW source modelling, data analysis, and astrophysical interpretation with a user-friendly tool for rapid background spectrum evaluation and easy integration of new models as our understanding of the GW universe expands.

\acknowledgements

We thank Patrick Meyers and Alan Weinstein for invaluable discussions and insight. %
We thank Thomas Callister for providing the population samples in~\cite{callister_zenodo}, and for carefully reading our work. %
We thank Nicholas Loutrel for consulting on waveform models and their features. %
AIR is supported by the European Union's Horizon 2020 research and innovation programme under the Marie Skłodowska-Curie grant agreement No 101064542, and acknowledges support from the NSF award PHY-1912594. JG is supported by NSF award No. 2207758. %
The authors are grateful for computational resources provided by the LIGO Laboratory and supported by National Science Foundation Grants PHY-0757058 and PHY-0823459. %
This material is based upon work supported by NSF's LIGO Laboratory which is a major facility fully funded by the National Science Foundation.
\appendix
\section{Deriving the Energy In GWs}\label{app:dEdf}

We derive here the energy spectrum $dE/df$ carried by gravitational waves, in vacuum. See, for example, \cite{Maggiore:2007ulw, Saffer18, Carroll04} %
We start by expanding the perturbed metric $g_{\mu\nu}$ to second order,
\begin{equation}
    g_{\mu \nu} = \eta_{\mu \nu} + h^{(1)}_{\mu \nu} + h^{(2)}_{\mu \nu},
\end{equation}
where $\eta_{\mu \nu}$ is the Minkowski flat metric and it is assumed that the perturbation $h^{(i)}_{\mu \nu}$ is $i$th order in some small parameter controlling the scale of $h_{\mu \nu}$. %
Substituting the above into the Einstein field equation gives
\begin{equation}
    G_{\mu \nu}\left[h^{(1)}\right] + G_{\mu \nu}\left[\left(h^{(1)}\right)^2\right] + G_{\mu \nu}\left[h^{(2)}\right]= 8 \pi G \tau_{\mu \nu}\,,
\end{equation}
Where the first term is the Einstein tensor linear in the first order perturbation, the second is the Einstein tensor terms quadratic in the first order perturbation, and the third term is linear in the second-order perturbation. %
In vacuum, $\tau_{\mu \nu} = 0$, and the solution for $h^{(1)}$ (i.e. the plane wave solution) reduces the first term to $ G_{\mu \nu}\left[h^{(1)}\right] = \Box h^{(1)}_{\mu \nu} = 0$, in the Lorenz gauge. %
We can therefore rearrange the above into a form that resembles the Einstein field equation,
\begin{equation}\label{eq:almosttheeinsteinfieldequation}
    G_{\mu \nu}\left[h^{(2)}\right] = -G_{\mu \nu}\left[\left(h^{(1)}\right)^2\right],
\end{equation}
where the first-order term $h^{(1)}_{\mu \nu}$ squared effectively forms a stress-energy (RHS) that sources the second order curvature (LHS). %
In this analogy to the RHS of the Einstein equation, we can define the effective stress energy (pseudo)-tensor of GWs \cite{Isaacson68}:
\begin{equation}\label{eq:stressenergyGW}
    \tau_{\mu \nu} \equiv -\frac{1}{8\pi G} G_{\mu \nu}\left[\left(h^{(1)}\right)^2\right].
\end{equation}

A nice feature here is that the LHS of Eq.~\ref{eq:almosttheeinsteinfieldequation} satisfies the contracted Bianchi identities and therefore the RHS is divergence-free and can be interpreted as conserving energy according to some observer.
Expanding out~\eqref{eq:stressenergyGW} gives
\begin{equation}
    \tau_{\mu \nu} = \frac{c^4}{32 \pi G} \left<\partial_{\mu} h_{\alpha \beta} \partial_{\nu} h^{\alpha \beta} \right>\,.
\end{equation}
The conservation law $\partial_\mu t^{\mu \nu} = 0$ implies
\begin{equation}\label{eq:conservation}
    \partial_0 \tau^{00} + \partial_i \tau^{i0} = 0\,,
\end{equation}
where $\tau^{00}$ can be interpreted as a volumetric energy density, such that the energy $E$ is defined as $E = \int d^3x \tau^{00}$ and therefore the associated power is
\begin{equation}
    \frac{dE}{dt} = \partial_0 \int d^3x \tau^{00}\,.
\end{equation}
Substituting into Eq.~\ref{eq:conservation} yields
\begin{equation}
    \frac{dE}{dt} + \int d^3x \partial_i \tau^{i0} = 0\,,
\end{equation}
which simplifies to
\begin{equation}
    \frac{dE}{dt} + \int d^3x \partial_z \tau^{z0} = 0
\end{equation}
for a wave moving along direction $\hat{z}$. %
We employ the divergence theorem to convert the volume integral into a surface integral,
\begin{equation}\label{eq:GWenflux}
    \frac{dE}{dt} + \hat{z} \int dA  \tau^{z0} = 0\,,
\end{equation}
which gives an expression for the energy flux (energy per unit time per unit area):
\begin{equation}
    \frac{dE}{dAdt} = -\tau^{z0} \hat{z}\equiv -\tau^{00} \hat{z}\,,
\end{equation}
as $\partial_0 h_{ij} = -\partial_z h_{ij} = -\partial^0 h_{ij}$ holds for a wave solution. %
Considering our gauge, the only surviving terms are those for $\mu, \nu = 1$ or $2$,
\begin{widetext}
\begin{equation}
    \tau^{00} = \frac{c^4}{32 \pi G}\left< \partial_0 h_{11} \partial_0 h^{11} + \partial_0 h_{12} \partial_0 h^{12} + \partial_0 h_{21} \partial_0 h^{21} + \partial_0 h_{22} \partial_0 h^{22} \right>\,,
\end{equation}
\end{widetext}
and substituting in the polarization components of the wave $h_{\mu \nu}$ in the TT gauge yields
\begin{figure*}
    \centering
    \includegraphics[width=0.9\textwidth]{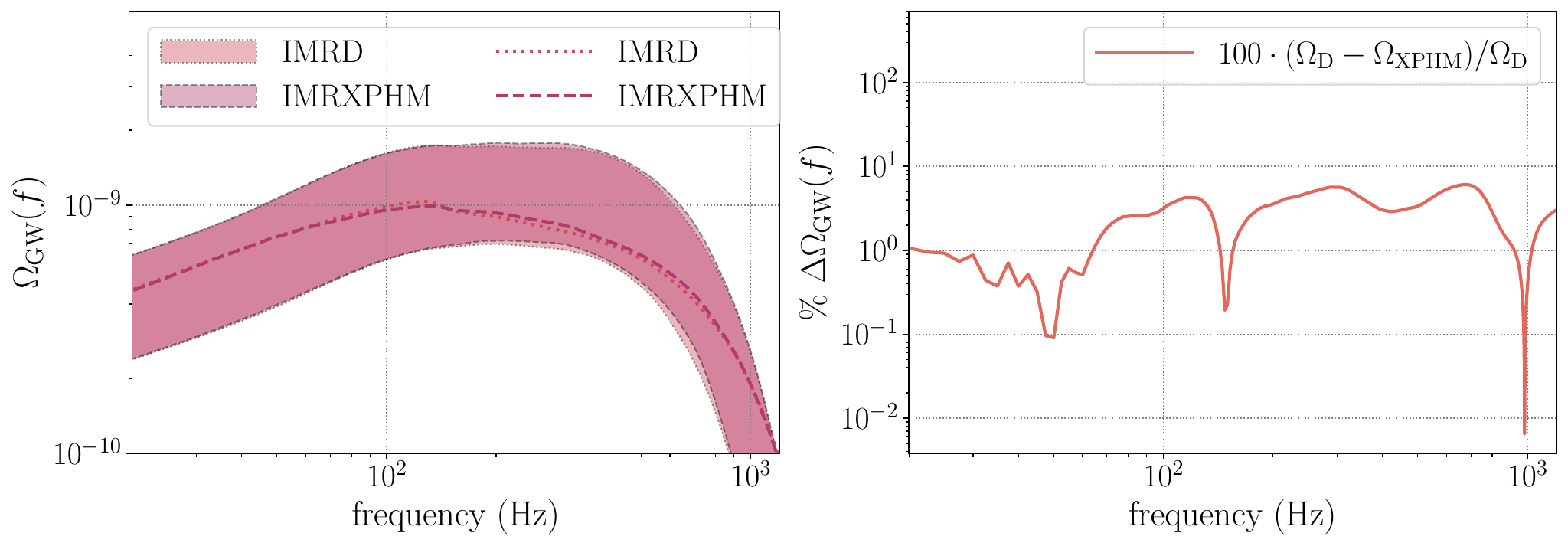}
    \caption{Impact of the inclusion of higher order modes in the waveform model employed to evaluate the $\OmegaGW$ spectrum. On the left: 95\% confidence on the spectrum including the uncertainty on the PLPP mass model and the local merger rate, assuming a fixed MD redshift evolution. On the right: percent difference \%$\Delta\OmegaGW(f)$ between $\OmegaGW$ spectra calculated using the same event samples, shown as dashed and dotted curves on the left panel. }
    \label{fig:waveforms_HOM}
\end{figure*}
\begin{equation}
    \tau^{00} = \frac{c^2}{16 \pi G} \left<|\dot{h}_{+}|^2 + |\dot{h}_{\cross}|^2 \right>\,.
\end{equation}
Solving for the energy flux of gravitational waves of Eq.~\eqref{eq:GWenflux} gives
\begin{equation}
    \left|\frac{dE}{dAdt}\right| = - \frac{c^3}{16 \pi G} \left<|\dot{h}_{+}|^2 + |\dot{h}_{\cross}|^2 \right>,
\end{equation}
where we have switched to the absolute value of this quantity with the understanding that GWs are removing energy from the system. %
The surface area energy density is then defined as
\begin{equation}\label{eq:surfaceareaenergydensity}
    \frac{dE}{dA} = \int dt \frac{c^3}{16 \pi G} \left< \dot{h}_{+}(t)^2 + \dot{h}_{\cross}(t)^2 \right>\,,
\end{equation}
To expand the above we recall that, for a plane wave,
\begin{equation}
    \dot{\tilde{h}}(f) = \int dt \dot{h}(t) e^{-i\omega t} = i\omega \int dt h(t) e^{-i\omega t} = i\omega \tilde{h}(f)\,,
\end{equation}
where in the final equivalence we have directly employed the definitaion of a Fourier transform. %
Recalling Parseval's theorem, 
we can write the surface area energy density in terms of 
the Fourier transform $\tilde{h}(f)$,
\begin{widetext}
\begin{equation}
    \frac{dE}{dA} = \int^{\infty}_{-\infty} df \frac{c^3 \omega^2}{16 \pi G} \left< \tilde{h}_{+}(f)^2 + \tilde{h}_{\cross}(f)^2 \right>= \frac{\pi c^3}{2 G} \int^{\infty}_{0} df f^2 \left< \tilde{h}_{+}(f)^2 + \tilde{h}_{\cross}(f)^2 \right>,
\end{equation}
\end{widetext}
where the integral is taken over the sphere surrounding the source. Note that $h_{+}$ and $h_{\cross}$ terms include dependence on inclination $\iota$ and reference phase $\phi_0$ (or, equivalently, the observer's position along the azimuth) and therefore must be included in the integral over the area. 

\section{Impact of higher order modes on the GWB spectrum}\label{app:HOMs}

For the sake of completeness, we append here findings on the impact on the $\OmegaGW$ spectrum due to the inclusion of higher order modes in the waveform model. %
Higher order modes are subdominant harmonics excited during GW emission, where the dominant harmonic is the $\ell=2$, $m=2$ mode~\cite{PhysRevLett.120.161102}. %
We compare the 95\% confidence bands shown in~\ref{sec:WFs} for the IMRPhenomD waveform model with bands obtained using the IMRPhenomXPHM approximant~\cite{Pratten:2020ceb}, which includes the $(\ell,\,|m|) =
(2,\,2),\,(2,\,1),\,(3,\,3),\,(3,\,2),\,(4,\,4)$ modes. %
As may be seen in Fig.~\ref{fig:waveforms_HOM}, for the population of binaries considered in~\ref{sec:WFs} which in particular is non-spinning and non-precessing, there are negligible differences between the use of IMRPhenomD and IMRPhenomXPHM. %
This is particularly evident when comparing the $\OmegaGW$ spectrum calculated using the same event samples employing the two waveforms, shown as dashed and dotted curves in the left panel of  Fig.~\ref{fig:waveforms_HOM}. The percent difference between these two curves is reported in the right panel of Fig.~\ref{fig:waveforms_HOM}, which remains consistently below 10\% across the spectrum and is under 3\% for frequencies below 100 Hz. 

\bibliography{bib}
\bibliographystyle{apsrev4-2}

\end{document}